\documentstyle[prl,aps,epsfig,twocolumn]{revtex}  

\begin{document}

\draft

\title{Wave Packet Echoes in the Motion of Trapped Atoms}

\author{F.B.J. Buchkremer, R. Dumke, H. Levsen, G. Birkl, and W. Ertmer}

\address{Institut f\"ur Quantenoptik, Universit\"at Hannover,
Welfengarten 1, D-30167 Hannover, Germany}

\date{\today}
\maketitle

\begin{abstract} 

We experimentally demonstrate and systematically study the stimulated revival
(echo) of motional wave packet oscillations. For this purpose, we prepare wave 
packets in an optical lattice by non-adiabatically shifting the potential and 
stimulate their reoccurence by a second shift after a variable time delay. This 
technique, analogous to spin echoes, enables one even in the presence of strong 
dephasing to determine the coherence time of the wave packets. We find 
that for strongly bound atoms it is comparable to the cooling time and much longer 
than the inverse of the photon scattering rate.

\end{abstract}

\pacs{32.80.Pj, 42.50.Vk}

\narrowtext


The process of decoherence, i.e. the collapse of superposition 
states due to the dissipative interaction with their environment is 
one of the basic concepts for our understanding of the connection between
classical and quantum physics.  
In order to study the effect of decoherence 
unambiguously, 
one has to be able to distinguish it from other, 
non-dissipative effects. The macroscopic (i.e. ensemble- or time-averaged) 
response of a 
quantum system prepared in a superposition state typically decays not 
only due to the loss of coherence (homogeneous decay) but also due to 
dephasing resulting from local variations in the evolution 
of the quantum system (inhomogeneous decay). In many cases
decoherence cannot be studied directly because the 
inhomogeneous decay
is by far the dominating process.

\begin{figure}
   \begin{center}
   \parbox{7.5cm}{
   \epsfxsize 7.5cm
   \epsfbox{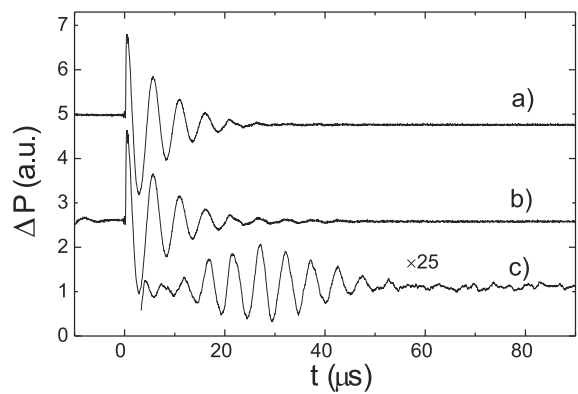}
}
   \end{center}
   
   \caption{Wave packet oscillations without (a) and with (b,c) stimulated
   revival (echo) ($U_{0}=831E_{R}$, $\delta$=- 7.8 $\Gamma$, 
   dz=0.10 $\lambda$).
   Depicted are                                
   the oscillations after the second shift (at t=0) for time delays 
   between the two shifts of $\Delta$t=108$\mu$s (a, reference curve, 
   complete decoherence before second shift) and 
   $\Delta$t=32$\mu$s (b, signal curve), showing additional oscillations
   around t=32$\mu$s. The difference between b) and a) (c, echo curve) 
   shows the net effect of the wave packet echo.}        
    \label{ECHOFIG1}
\end{figure}

This limitation has been overcome in a famous series of experiments 
by introducing the techniques of spin echo 
for nuclear 
magnetic resonance (NMR) and photon echo
for optical resonance, respectively \cite{HAHN,KURNIT,ALLEN}. 
These techniques are based on the observation that 
inhomogeneous decay due to dephasing is a reversible process. Thus, by 
appropriately 
modifying superposition states at a time $\Delta$t 
after their preparation, 
the 
dephasing can be partially or fully reversed and a stimulated 
macroscopic response (echo)
is induced at 2$\Delta$t. This effect enables one to 
measure
the coherence time even in the presence of strong dephasing. We have, for the first time, 
applied this method to the investigation of the decoherence of motional wave packets of trapped
atoms (Fig. \ref{ECHOFIG1}).
The method can be used independent of the specific experimental realization of the confining
potential (e.g. a single dipole potential, periodic dipole potentials, magnetic trapping 
potentials, inhomogeneous arrays of atom traps, etc.).

The specific system investigated here consists of motional wave packets
of neutral atoms  
in a one-dimensional optical lattice.
Optical lattices are periodic dipole
potentials for atoms created
by the interference of multiple laser beams \cite{JESSENMEACHERGUIDONI}. 
Atoms can be 
trapped and cooled at the 
potential minima (mean position spread $z_{rms}$=$\lambda$/18 \cite{GATZKE}).  
In optical lattices symmetrically and asymmetrically 
oscillating motional wave packets can be induced by non-adiabatically changing the lattice 
potential \cite{KOZUMA,RAITHEL1,GORLITZ,RAITHEL2,RUDYEJNISMAN,RAITHEL3}. 
Quantum mechanically, the original atomic wave function is projected 
onto a 
coherent 
superposition of the
eigenstates of the new potential and the quantum
interference of the contributions from different eigenstates results in 
a wave packet oscillation   
(see Fig. \ref{ECHOFIG1}(a)).

In dipole potentials, the macroscopic oscillation signal decays because of 
decoherence due to the 
spontaneous scattering of photons and because of dephasing due to the 
anharmonicity of the potential wells and spatial variations 
of the potential depth. Typically, 
the effect of dephasing is dominating decoherence 
\cite{KOZUMA,RAITHEL1,GORLITZ,RAITHEL2,RUDYEJNISMAN,RAITHEL3,MORSCH,MORINAGA},
so that a direct determination of the coherence time is not possible. 
Here, we show how these 
limitations can be overcome:
For the case of 
symmetrical oscillations,
Bulatov {\it et al.} \cite{BULATOV} have 
recently proposed and numerically simulated an echo-mechanism to reverse  
the effect of dephasing and stimulate the revival of the
wave packet oscillations by means of two successive non-adiabatic changes in the
depth of the lattice potential. 
With our work we

extend their proposal to the case of 
asymmetrical wave packet oscillations, present the first experimental 
observation 
of the echo effect, and apply it to determine the coherence time of
the motional wave packets.

In our experiment, we chirp-slow rubidium atoms ($^{85}$Rb) from a thermal 
beam and trap and cool them in a magneto-optical trap (MOT) giving a sample of  
approximately $10^7$ atoms with a central density of $10^{9}$ atoms/$cm^3$. 
After the loading
phase the magnetic field of the MOT is switched off and the intensity 
of the trapping beams is reduced to achieve optimal cooling in a 
3-dimensional molasses. The MOT-laser beams are then turned off and two 
lattice beams are switched on, forming a 1-dimensional
lin${\perp}$lin optical lattice \cite{JESSENMEACHERGUIDONI}. The lattice 
beams have 
intensities of up 
to $I = 60$ $mW/cm^2$, 
and detunings $\delta$ of 2 to 10 natural linewidths ($\Gamma/2\pi=5.89 MHz$) 
below 
the $5S_{1/2} (F=3) \rightarrow 5P_{3/2} (F'=4)$ transition at 
$\lambda=780 nm$. 
The beam waist of about 2.75mm ($1/e^2$ radius)
is large compared to the $1/e^2$ radius of the atomic cloud 
of 1.52 mm.

After an initial cooling phase of 1 to 2 ms in the lattice, which
localizes  
the atoms at the center of the potential wells, 
we non-adiabatically change the 
relative 
phase between 
the two lattice beams with 
an electrooptic phase shifter \cite{RAITHEL3} (1/e switching time 
of 0.4 $\mu$s). This causes a 
translation of the lattice 
by a controllable 
amount $dz$ ($0<dz<\lambda/4$)
and induces asymmetrical coherent-state-like motional wave packets.
We observe the wave packet oscillations by measuring the 
photon redistribution-induced
power difference $\Delta$P(t) between the two lattice beams 
\cite{RAITHEL3,FOOTNOTE1}. No repumping light is present during 
the wave packet evolution.
Fig. \ref{ECHOFIG1}(a) shows a typical example of a wave packet 
oscillation. Clearly 
visible are about 5 oscillations with a period of $(5.2\pm0.1)\mu$s. The oscillation
signal is damped with a decay time of $\tau_{1}=(7\pm1)\mu$s 
(exponential fit).

\begin{figure}
   \begin{center}
   \parbox{7.5cm}{
   \epsfxsize 7.5cm
   \epsfbox{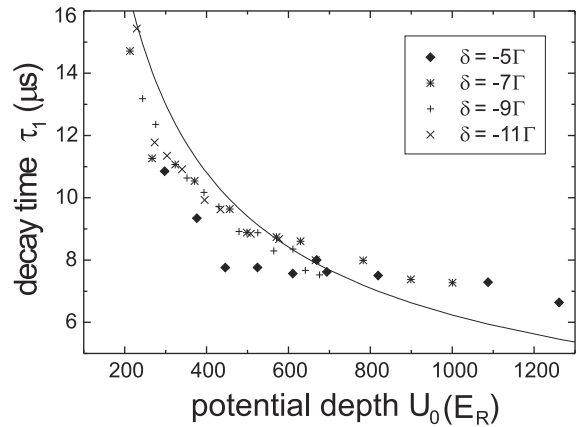}
}
   \end{center}
   
   \caption{Decay times $\tau_{1}$ of asymmetrical wave packet 
   oscillations as a 
   function of the potential depth $U_{0}$ for the 
   indicated values of the detuning $\delta$. 
   The solid line shows a calculation of the decay time for 
   anharmonicity induced dephasing as the decay mechanism.}         
   \label{ECHOFIG2}
\end{figure}

Fig. \ref{ECHOFIG2} shows the decay time 
$\tau_{1}$ (relative
uncertainty $\le$ 20\%) of asymmetrical wave packet oscillations 
as a function of the 
potential depth $U_{0}$ for 
various detunings $\delta$. 
$U_{0}$ is calculated from the measured oscillation 
frequency $\omega_{osc}$ according to 
$U_{0}/E_{R}=(1/(0.86)^2)(\omega_{osc}/2\omega_{R})^2$ 
taking into account the
anharmonicity of the potential wells \cite{GATZKE,RAITHEL2}.  
The data show that the decay is not due to decoherence 
caused by spontaneous 
scattering, for in that case
the decay time $\tau_{1}$ should be proportional to the inverse of the 
photon scattering rate, $\tau_{sc} = 1/\Gamma'= 1/(U_{0}\Gamma/(\hbar|\delta|)$.
For different detunings this
should result in different 
decay times $\tau_{1}$ for the same $U_{0}$, which is 
not observed for $|$$\delta$$|$ $\ge$ 7$\Gamma$. 
For a decay caused by anharmonicity on the other hand, 
the decay time should be proportional to the 
inverse of the 
mean spread $\Delta\omega_{osc}$
of the occuring
oscillation frequencies,  
which can be approximated 
by the mean shift of the oscillation frequencies from the harmonic frequency minus
the common shift of $\omega_{R}$,
i.e. $\Delta\omega_{osc}\approx 0.14 \omega_{osc}-\omega_{R}$ \cite{GATZKE,RAITHEL2}.
The line in Fig. \ref{ECHOFIG2} is a plot of 
$1/\Delta\omega_{osc}$ as a function of
$U_{0}$.
Within the 20\%

uncertainty of our data
the decay times $\tau_{1}$ are in agreement with this
calculation.
This shows
that the decay of the wave packet
oscillations is dominated by dephasing
\cite{FOOTNOTE1a}.

The investigation of decoherence, on the other hand, becomes possible
by non-adiabatically shifting the lattice 
back to its initial position after a 
variable delay $\Delta$t. This second 
translation leads to a stimulated revival (echo) of the 
oscillations with a maximum amplitude at about 2$\Delta$t \cite{BULATOV}
if the coherence of the wave packets induced by the first shift 
still persists - at least partially - at 2$\Delta$t.  

Fig. \ref{ECHOFIG1} shows the first experimental 
demonstration of this echo effect \cite{FOOTNOTE2}. 
Fig. \ref{ECHOFIG1} (a) serves as a reference curve showing 
wave packet oscillations for atoms that move fully incoherently at the 
time of the shift. For the 
purpose of reducing systematic uncertainties, this curve  
also has been recorded after two translations of the 
lattice. However, 
here the long delay time $\Delta$t=108$\mu$s  
guarantees a complete loss of coherence before the second shift. 
Fig.\ref{ECHOFIG1} (b) depicts the wave 
packet oscillations after the second shift with 
$\Delta$t=32$\mu$s.
The curve shows additional oscillations at about t=32$\mu$s.  
Fig.\ref{ECHOFIG1} (c) presents the magnified difference of curves (b) 
and (a). Clearly visible is the reocurrence of wave packet oscillation, 
i.e. echo,  
centered around a time close to t=32$\mu$s. This time corresponds to a total 
time 
of $t_{total}$=32$\mu$s+$\Delta$t=64$\mu$s=2$\Delta$t after the first shift, 
as predicted.

\begin{figure}
   \begin{center}
   \parbox{7.5cm}{
   \epsfxsize 7.5cm
   \epsfbox{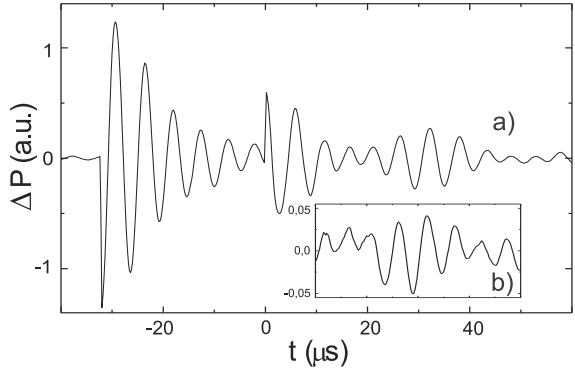}
}
   \end{center}
   
   \caption{
   QMCWF simulations of the echo experiment ($I=56.5 mW/cm^2$ and 
   $\delta$=- 7.8 $\Gamma$ ($\Rightarrow$ $U_{0}=823E_{R}$), dz=0.064 $\lambda$, $\Delta$t=32$\mu$s
   (potential shifts at $t=-32\mu$s and $t=0$)).
   (a) With a reduced rate of spontaneous scattering processes ($\Gamma'\rightarrow\Gamma'/1000$)
   a strong wave packet echo is visible. (b)
   With the regular rate of spontaneous scattering the echo is still visible 
   but its amplitude is significantly reduced.}         
   \label{ECHOFIG3}
\end{figure}

In order to gain a qualitative and quantitative understanding of the echo mechanism,
we have performed a full quantum Monte-Carlo wave-function 
simulation 
(QMCWF) \cite{MonteCarlo} of the echo experiment taking into account the full coherent and dissipative dynamics 
as well as all internal ground states while adiabatically eliminating the excited states.
Fig. \ref{ECHOFIG3} shows the result for the experimental conditions
of Fig. \ref{ECHOFIG1}. In Fig. \ref{ECHOFIG3} (a) the spontaneous
scattering rate is reduced ($\Gamma'\rightarrow\Gamma'/1000$) in order to

emphasize the coherent dynamics which gives rise 
to the echo
mechanism. Clearly visible 
is the occurence of the wave packet
echo at t = 32 $\mu$s.
                                
The underlying physics of the echo mechanism can be understood in terms of 
the coherent evolution of the contributions
of different eigenstates to the wave packet: 
The first translation creates a coherent superposition of
eigenstates of the translated anharmonic trapping potential which are all in phase 
(with phase = 0 per definition). 
The phases of different eigenstates evolve with different frequencies in time
which causes the decay of the oscillation signal. Shifting the lattice back 
after $\Delta$t creates a new superposition state.
Due to the asymmetry of the translation,
eigenstates with phases close to odd multiples of $\pi$ 
give the strongest and 
eigenstates with phases close to even multiples of $\pi$
give the weakest contributions to the new superposition state
which leads to
new wave packet oscillations which again dephase in time.
However, eigenstates that were in phase at t=$\Delta$t are again in phase at t=2$\Delta$t, 
so that the strongest contributions again realign at t=2$\Delta$t which results in a partial 
revival of the oscillations.
Thus, the phase-dependent selection of the strength of eigenstate contributions at the 
second
shift leads to the wave packet 
echo.
 
Spontaneous scattering leads to an incoherent evolution of the superposition
state by randomizing the phases of the eigenstates 
and thus causes a decrease in the echo amplitude. This is clearly 
observed in our wave 
function simulation for a non-reduced spontaneous scattering rate (Fig. \ref{ECHOFIG3} (b)).
The wave packet echo is still visible but its amplitude is reduced to 
a value that is consistent with the one obtained in the experiment (Fig. \ref{ECHOFIG1}).  

In order to experimentally study additional effects of dephasing, we have narrowed the 
transverse
intensity profile of our lattice
beams causing different parts of the atom
cloud to experience different potential depths $U_0$.

\begin{figure}
   \begin{center}
   \parbox{7.5cm}{
   \epsfxsize 7.5cm
   \epsfbox{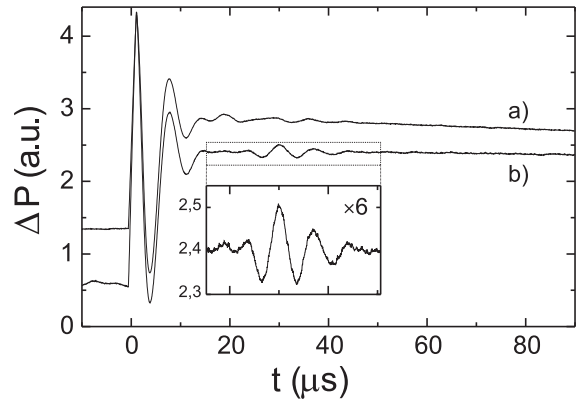}
}
   \end{center}
   
   \caption{Wave packet oscillations and echo for
   increased dephasing ($U_{0}=368E_{R}$, $\delta$=-8 $\Gamma$, 
   dz=0.11 $\lambda$). The reference curve (a, $\Delta$t=132$\mu$s)
   depicts the faster signal decay. The signal curve 
   (b, $\Delta$t=30$\mu$s) shows a free standing echo.}        
   \label{ECHOFIG4}
\end{figure}
 
The almost free-standing echo in the signal curve
of Fig. \ref{ECHOFIG4}
experimentally
demonstrates,
that the
method of wave packet echoes can also be applied to other systems 
where the spatial 
variation in the coherent
dynamics is the dominating effect for dephasing
as e.g. in 
anharmonic dipole traps based on focused laser beams, inhomogeneous arrays of atom traps, or 
magnetic quadrupole traps, etc.
(see e.g. \cite{MORINAGA}).

\begin{figure}[b]
   \begin{center}
   \parbox{7.5cm}{
   \epsfxsize 7.5cm
   \epsfbox{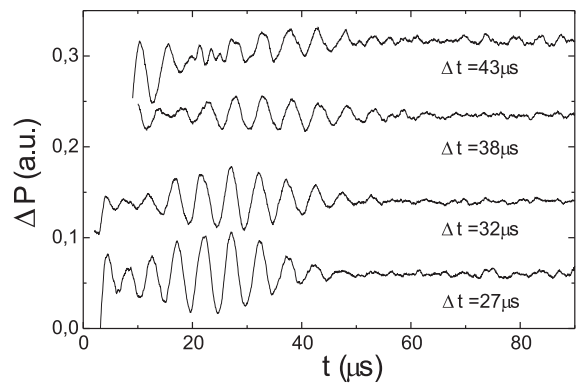}
}
   \end{center}
   
   \caption{Increase in occurence time and decrease in amplitude 
   of the echoes for increasing delay 
   $\Delta$t between translations of the trapping potential (parameters 
   as in Fig. \ref{ECHOFIG1}).}     
   \label{ECHOFIG5}
\end{figure}

The physics behind the echo effect predicts that the time at which the
echo occurs can be varied by changing the time delay 
$\Delta$t between lattice shifts. Fig. \ref{ECHOFIG5} 
shows echo curves (analogous to Fig.\ref{ECHOFIG1} (c))
for the indicated values of the time delay $\Delta$t.
For each curve, t=0 corresponds to the second lattice shift. The total time 
after the first shift is given by $t_{total}$=t+$\Delta$t.
Our measurements 
confirm
that
the echo moves to larger t with increasing $\Delta$t and
has
its maximum value at approximately $t_{total}$=2$\Delta$t 
\cite{FOOTNOTE3}.

\begin{figure}
   \begin{center}
   \parbox{7.5cm}{
   \epsfxsize 7.5cm
   \epsfbox{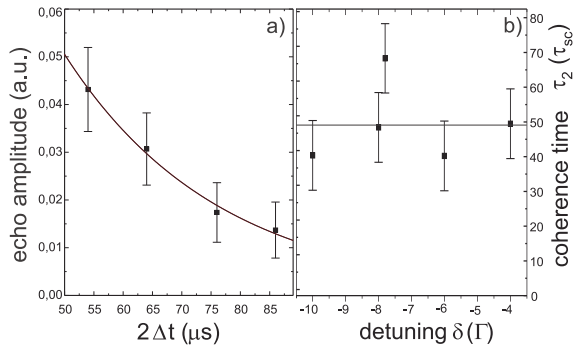}
}
   \end{center}
   
   \caption{(a) Dependence of the 
   echo amplitude on the echo time 2$\Delta$t for the data of 
   Fig. \ref{ECHOFIG5}. The line is an 
   exponential fit to the data, giving a coherence time of $\tau_2$=(27$\pm$3) $\mu$s. 
   (b) Coherence time in units of the photon scattering time 
   $\tau_{sc} = 1/\Gamma'$ as a function of the lattice detuning $\delta$.} 
   \label{ECHOFIG6}
\end{figure}

We can use the echo technique to determine the coherence time $\tau_2$ of the 
atomic motion by 
measuring the 
echo 
amplitude as a
function of the
echo time 2$\Delta$t. Fig. \ref{ECHOFIG6} (a) shows
this for
the data of Fig. \ref{ECHOFIG5}. With 
increasing
$\Delta$t 

the echo
amplitude decreases exponentially 
as expected for a coherence
loss being induced by spontaneous scattering
processes occuring at a constant rate to $\Gamma'$.
An exponential fit to the data (line
in 
Fig. \ref{ECHOFIG6} (a)) gives a 1/e decay time of 
$\tau_2$=(27$\pm$3) $\mu$s
which is 
significantly longer than the observed dephasing time of 7 $\mu s$. This 
shows that the coherent motion still persists for 
times at which the macroscopic oscillation signal has already disappeared.

The echo technique even enables a quantitative determination of the constant of 
proportionality between the coherence 
time $\tau_2$ and the photon scattering time $\tau_{sc}=1/\Gamma'$.
For the data of
Fig. \ref{ECHOFIG6} (a), $\tau_{sc}$ = (0.40$\pm$0.04) $\mu$s and we
find that during the coherence time one atom scatters  
68$\pm$10 photons. In a series of measurements similar to 
Fig. \ref{ECHOFIG5} for -10$\Gamma\le\delta\le-4\Gamma$ we observe 
coherence times in the range from 40 to 68 scattering times $\tau_{sc}$
with an
average value of $\tau_2 = (49 \pm 7) \tau_{sc}$ independent of the detuning 
$\delta$ (Fig. \ref{ECHOFIG6} (b)). This proves 
that the coherence time $\tau_2$ is proportional to the scattering time $\tau_{sc}$ 
and that for the decay of motional coherence a 
large
number of photons has to be scattered spontaneously.  

The quantitative determination of the coherence time enables us to prove an important
prediction given for the dissipative coupling of the motion of atoms in (nearly) harmonic
potentials to their environment: 
It has been pointed out that for (nearly) harmonic oscillators
the coherence time $\tau_2$ should be twice as long as
the damping time of the
oscillator's energy, i.e. the cooling time $\tau_{cool}$ of the atoms 
\cite{PHILLIPS,CIRAC,COHEN}. 
In a recent experiment, $\tau_{cool}$ in one-dimensional
optical lattices was found to be the time
to spontaneously 
scatter 30 photons: $\tau_{cool}=30\tau_{sc}$ 
\cite{RAITHEL1}. 
Our direct measurement of the
coherence time $\tau_2 = (49 \pm 7) \tau_{sc}$
is the first experimental confirmation
of the predicted relation between the coherence time and the cooling time.
We expect the relation $\tau_2 =2\tau_{cool}$ 
to hold universally for (nearly) harmonic trapping potentials in which the motion of atoms is coupled 
to a dissipative reservoir \cite{PHILLIPS}.

In summary, we have experimentally demonstrated and systematically investigated a new echo technique
to access the coherent and
dissipative dynamics
of atomic wave packet
oscillations in trapping potentials. With this technique, we have 
measured the coherence time of wave 
packet oscillations in strongly confining dipole potentials, in a regime, which was previously
inaccesible. We showed that the coherence time is directly connected to the energy dissipation 
time. We also 
demonstrated that the echo technique works for atomic motion exhibiting various causes of 
dephasing which shows that it can be used to access the external dynamics in a broad range of different 
atom traps 
or inhomogeneous arrays of atom traps.

We thank A. Pahl and M. Wilken for their support in the early stages of 
the project. This work is supported by the SFB 407 of the {\it Deutsche 
Forschungsgemeinschaft}.



\begin{references}

\bibitem{HAHN} E.L. Hahn,
Phys. Rev. {\bf 80}, 580 (1950).

\bibitem{KURNIT} N.A. Kurnit, I.D. Abella, and S.R. Hartmann,
Phys. Rev. Lett. {\bf 13}, 567 (1964).

\bibitem{ALLEN} L. Allen and J.H. Eberly,
\it Optical Resonance and Two-Level Atoms \rm (Dover Publications, 
New York, 1987). 

\bibitem{JESSENMEACHERGUIDONI} P.S. Jessen and I.H. Deutsch, Adv. At. Mol. Opt.
Phys. {\bf 37}, 95 (1996); D.R. Meacher, Cont. Phys. {\bf 39}, 329 (1998);
L. Guidoni and P. Verkerk, J. Opt. 
B {\bf 1}, R23 (1999).

\bibitem{GATZKE} M. Gatzke, G. Birkl, P.S. Jessen, A. Kastberg, 
S.L. Rolston, and W.D. Phillips,  
Phys. Rev. A {\bf 55}, R3987 (1997).
                                                                    
\bibitem{KOZUMA} M. Kozuma, N. Nakagawa, W. Jhe, and M. Ohtsu,
Phys. Rev. Lett. {\bf 76}, 2428 (1996).

\bibitem{RAITHEL1} G. Raithel, G. Birkl, A. Kastberg, W.D. Phillips, 
and S.L. Rolston, Phys. Rev. Lett. {\bf 78}, 630 (1997).

\bibitem{GORLITZ} A. G\"orlitz, M. Weidem\"uller, T.W. H\"ansch,
and A. Hemmerich, Phys. Rev. Lett. {\bf 78}, 2096 (1997).

\bibitem{RAITHEL2} G. Raithel, G. Birkl, 
W.D. Phillips, and S.L. Rolston,
Phys. Rev. Lett. {\bf 78}, 2928 (1997).

\bibitem{RUDYEJNISMAN} P. Rudy, R. Ejnisman, and N.P. Bigelow, 
Phys. Rev. Lett. {\bf 78}, 4906 (1997),
R. Ejnisman, P. Rudy, H. Pu, and N.P. Bigelow, 
Phys. Rev. A. {\bf 56}, 4331 (1997).

\bibitem{RAITHEL3} G. Raithel, W.D. Phillips, and S.L. Rolston,
Phys. Rev. Lett. {\bf 81}, 3615 (1998).

\bibitem{MORSCH} O. Morsch, P.H. Jones, and D.R. Meacher, Phys. Rev. A 
{\bf 61}, 023410 (2000).

\bibitem{MORINAGA} M. Morinaga, I. Bouchoule, J.-C. Karam, 
and C. Salomon, Phys. Rev. Lett. {\bf 83}, 4037 (1999).

\bibitem{BULATOV} A. Bulatov, A. Kuklov, B.E. Vugmeister, and H. Rabitz,
Phys. Rev. A {\bf 57}, 3788 (1998).
                           
\bibitem{FOOTNOTE1} The maximum power difference $\Delta$P 
is about 1\% 
of the total lattice power for all data presented here.           
           
\bibitem{FOOTNOTE1a} The remaining deviations
may be due to the additional influence of decoherence and local variations in $U_{0}$. 

\bibitem{FOOTNOTE2} In order to compensate fluctuations in 
atom number 
the initial amplitudes of all wave packet oscillations are normalized. 
For clarity some curves are offset vertically. 

\bibitem{MonteCarlo} J. Dalibard, Y. Castin, and K. M{\o}lmer, 
Phys. Rev. Lett. {\bf 68}, 580 (1992),
P. Marte, R. Dum, R. Ta\H{\i}eb, and P. Zoller, Phys. Rev. A {\bf 47}, 1378 (1993),
P. Marte, R. Dum, R. Ta\H{\i}eb, P.D. Lett, and P. Zoller, Phys. Rev. Lett. {\bf 71},
1335 (1993).

\bibitem{FOOTNOTE3} The fact that the maxima occur at times slightly 
earlier than 2$\Delta$t is caused by decoherence, which decreases 
the signal strength during the echo, thus shifting the 
apparent echo maximum to earlier times.

\bibitem{PHILLIPS} W.D. Phillips and C.I. Westbrook,
Phys. Rev. Lett. {\bf 78}, 2676 (1997).

\bibitem{CIRAC} J.I. Cirac, R. Blatt, A.S. Parkins, 
and P. Zoller, Phys. Rev. A. {\bf 48}, 2169 (1993).

\bibitem{COHEN} C. Cohen-Tannoudji, J. Dupont-Roc, and
G. Grynberg, {\it Atom-Photon Interactions} (Wiley, New York, 1992).







\end{references}
\end{document}